\documentclass{emulateapj}

\usepackage{psfig}
\usepackage{natbib}
\newcommand\oneprime{\mbox{$^{\prime}$}}%

\shorttitle{The Fourth Positive System of CO in Comet Spectra}
\shortauthors{Lupu et al.}

\begin{document}


\title{The Fourth Positive System of Carbon Monoxide in the Hubble Space Telescope Spectra of Comets}


\author{Roxana E. Lupu, Paul D. Feldman}
\affil{Department of Physics and Astronomy, Johns Hopkins University,
3400 N. Charles Street, Baltimore, MD 21218}
\email{roxana@pha.jhu.edu}

\author{Harold A. Weaver}
\affil{Space Department, Johns Hopkins
University Applied Physics Laboratory, \\
11100 Johns Hopkins Road, Laurel, MD 20723}

\and

\author{Gian-Paolo Tozzi}
\affil{INAF - Osservatorio Astrofisico di Arcetri, Largo E. Fermi 5,
I-50125 Florence, Italy}
    




\begin{abstract}

The rich structure of the $A^{1}\Pi-X^{1}\Sigma^{+}$ system of carbon
monoxide accounts for many of the spectral features seen in long slit
$HST$-STIS observations of comets 153P/Ikeya-Zhang, 
C/2001 Q4 (NEAT), 
and C/2000 WM$_{1}$ (LINEAR), 
as well as in the $HST$-GHRS spectrum of comet C/1996 B2 (Hyakutake). 
A detailed CO fluorescence model is developed to derive the CO
abundances in these comets by simultaneously fitting all of the observed
$A-X$ bands. The model includes the latest values for the oscillator
strengths and state parameters, and accounts for optical depth effects
due to line overlap and self-absorption. A complete fitting of the CO
spectral features using this model leads to the first identification of a
molecular hydrogen line pumped by solar \ion{H}{1} Lyman-$\beta$ longward
of 1200~\AA\ in the spectrum of comet 153P/Ikeya-Zhang. Pumping by
strong solar lines also plays an important role in CO fluorescence, as
shown by the detection in the spectrum of comet C/1996 B2 (Hyakutake)
of bands from the (14-v\arcsec) and (9-v\arcsec) progressions pumped by
solar \ion{H}{1} Lyman-$\alpha$ and \ion{O}{1} $\lambda$1302,
respectively.  Using spectra extracted at increasing distances from the
comet nucleus or averaging over increasing effective apertures, the
model fits yield radial profiles of CO column density that are
consistent with a predominantly native source for all the comets
observed by STIS. The derived CO production rates are $1.54 \pm 0.09
\times 10^{28}$~molecules~s$^{-1}$ for 153P/Ikeya-Zhang, $1.76 \pm 0.16
\times 10^{28}$~molecules~s$^{-1}$ for C/2001 Q4 (NEAT), and
$3.56 \pm 0.20 \times 10^{26}$~molecules~s$^{-1}$ for C/2000 WM$_{1}$
(LINEAR). In the absence of spatial information for comet C/1996 B2
(Hyakutake), we estimate a CO production rate of $4.97 \pm 0.07 \times
10^{28}$~molecules~s$^{-1}$, assuming an entirely native source. The CO
abundances relative to water in these comets span a wide range, from
$0.44 \pm0.03$\% for C/2000 WM$_{1}$ (LINEAR), $7.2 \pm 0.4$\% for
153P/Ikeya-Zhang, $8.8 \pm 0.8$\% for C/2001 Q4 (NEAT) to $20.9 \pm
0.3$\% for C/1996 B2 (Hyakutake).  For comets C/2000 WM$_{1}$ and
C/2001 Q4 we can compare these results with those derived from nearly
simultaneous observations by the {\it Far Ultraviolet Spectroscopic
Explorer}.

\end{abstract}



\section{INTRODUCTION}

Cometary ices have undergone little or no processing since their formation in the solar nebula and thus they represent an important clue in understanding the conditions in the early Solar System. However, this interpretation is complicated by the radial mixings in the protoplanetary disk and the dynamical evolution of comets, expelled to the outer Solar System from their original formation site. No pattern for the cometary composition and activity has emerged from the objects studied to date \citep{biver02,crovisier07} and a larger sample is needed. The composition of cometary ices has been studied through spectroscopic observations or {\it in situ} measurements. In the analysis of cometary spectra it is important to establish whether the observed chemical species originate in the nucleus (native source) or in the coma (extended source). For this purpose, spatially-resolved observations are needed to derive the radial composition of the gas outflow. 

Carbon monoxide (CO) is one of the most abundant molecules, whose mixing ratio relative to water varies greatly among comets \citep{biver06,cometsii}. As a highly volatile compound, with a sublimation temperature of 24~K, CO ice is thought to be a sensitive tracer of the temperature in the environment in which the comets formed. However, its origin as a native compound or a daughter product, and the correlation between the two sources is still not understood. Infrared observations of the CO ro-vibrational lines conclude that the CO source was mainly native in comet C/1996~B2~(Hyakutake) \citep{disanti03} and significantly extended for comet C/1995~O1~(Hale-Bopp) \citep{disanti01}. This paper complements previous findings with the first CO radial profiles constructed from ultraviolet spectra. Care must be taken in interpreting the brightness profiles, since saturation close to the nucleus could mimic the presence of an extended source. A comprehensive fluorescence model is developed to derive reliable CO column densities and estimate CO production rates from cometary spectra obtained by the Space Telescope Imaging Spectrograph (STIS) and the Goddard High Resolution Spectrograph (GHRS) on the {\it Hubble Space Telescope} ({\it HST}). Using the long-slit capabilities of the STIS instrument, this paper offers evidence that the native source of CO is dominant in three long period comets.

\begin{deluxetable*}{cccccccc}
\tabletypesize{\footnotesize}
\tablecaption{Comet observations. \label{table1}}
\tablewidth{0pt}
\tablehead{
\colhead{Dataset} & \colhead{Date \& Time}   & \colhead{Exposure Time}   &
\colhead{r$_{h}$} & \colhead{ v$_{h}$}  & 
\colhead{$\Delta$} & \colhead{Instrument}& \colhead{Mode}\\ 
	& (UT) & (s) & (AU) & (km~s$^{-1}$) & (AU) & & }
\startdata
\sidehead{C/1996 B2 (Hyakutake)}
Z35FN602T & 1996-04-01  13:09:00 & 1305.600 & 0.885 & --38.51 & 0.259 & $HST$-GHRS & G140L;2.0\\
Z35FN604T & 1996-04-01  13:39:00 &  217.600 & 0.885 & --38.51 & 0.260 & $HST$-GHRS & G140L;2.0\\
Z35FN702T & 1996-04-01  14:44:00 & 1305.600 & 0.884 & --38.53 & 0.262 & $HST$-GHRS & G140L;2.0\\
Z35FN704T & 1996-04-01  15:14:00 &  217.600 & 0.883 & --38.53 & 0.262 & $HST$-GHRS & G140L;2.0\\
\tableline
\sidehead{C/2000 WM$_{1}$ (LINEAR)}
B0500301000 & 2001-12-07  08:50:00 &  &  &  &  &  & \\
to B0502401000 & 2001-12-10  06:48:00 & 36,467 & 1.120 & --28.30 & 0.340 & $FUSE$ & LWRS\\
O6GR12010 & 2001-12-09  21:33:00 & 1800.194 & 1.085 & --28.26 & 0.357 & $HST$-STIS & G140L;52$\times$0.2\\
O6GR03010 & 2001-12-09  23:09:10 & 1440.197 & 1.084 & --28.26 & 0.358 & $HST$-STIS & G140L;52$\times$0.2\\
O6GR11010 & 2001-12-10  00:45:00 & 1800.198 & 1.083 & --28.26 & 0.359 & $HST$-STIS & G140L;52$\times$0.2\\
\tableline
\sidehead{153P/Ikeya-Zhang}
O8FY01010 & 2002-04-20  07:28:00 & 1800.199 & 0.887 & 29.07 & 0.426 & $HST$-STIS & G140L;52$\times$0.2\\
O8FY02010 & 2002-04-20  10:40:00 & 1800.197 & 0.889 & 29.08 & 0.426 & $HST$-STIS & G140L;52$\times$0.2\\
O8FY03010 & 2002-04-21  07:30:00 & 1440.197 & 0.904 & 29.15 & 0.422 & $HST$-STIS & G140L;52$\times$0.2\\
\tableline
\sidehead{C/2001 Q4 (NEAT)}
E1390101000 & 2004-04-24  00:39:00 &  &  &  &  &  & \\
to E1390501000 & 2004-04-24  23:09:00 & 68,282 & 1.030 & --10.80 & 0.510 & $FUSE$ & LWRS\\
O8VK04010 & 2004-04-25  20:03:00 & 1800.199 & 1.024 & --10.26 & 0.473 & $HST$-STIS & G140L;52$\times$0.2\\
O8VK01010 & 2004-04-26  00:49:00 & 1683.008 & 1.023 & --10.18 & 0.468 & $HST$-STIS & G140L;52$\times$0.2\\
O8VK07010 & 2004-04-30  00:51:00 & 1800.200 & 1.002 &  --8.37 & 0.386 & $HST$-STIS & G140L;52$\times$0.2\\
\enddata
\end{deluxetable*}

CO ultraviolet fluorescence in cometary spectra was first detected
during sounding rocket observations of comet West 
\citep[C/1975 V1,][]{feld76} and has been subsequently observed by $IUE$
\citep{tozzi98}, $HST$ \citep{Weaver:1998}, $FUSE$
\citep{feld02b}, the Hopkins Ultraviolet Telescope on
the {\it Astro-1} Space Shuttle mission \citep{feld91}, and rockets
\citep{woods87,sahnow93,mcp99}. The most important spectral features
of CO in the ultraviolet (UV) are its electronic transitions
belonging to the $A-X$, $B-X$ and $C-X$ systems, or to
the forbidden Cameron bands. Fluorescence is the main emission
mechanism for the $A^{1}\Pi-X^{1}\Sigma^{+}$ system (1300~--~1900~\AA),
$C^{1}\Sigma^{+}-X^{1}\Sigma^{+}$ system (0--0 band at
1087.9~\AA), and $B^{1}\Sigma^{+}-X^{1}\Sigma^{+}$ system (0--0
band at 1150.5~\AA). The Cameron bands, $a^{3}\Pi-X^{1}\Sigma^{+}$
(1900~--~2800~\AA), are mainly excited by electron impact and
photodissociation of CO$_{2}$ \citep{Weaver:1994}.  Although
observations of the $A-X$ or Fourth Positive Group of CO in the UV
spectra of comets have a long history, their interpretation has been
difficult compared to that of the $B-X$ and $C-X$ systems at shorter
wavelengths that have been observed more recently.

The spatial information offered by the high resolution long slit
$HST$-STIS spectra of comets 153P/Ikeya-Zhang~(C/2002~C1),
C/2001~Q4~(NEAT), and C/2000~WM$_{1}$~(LINEAR) shows
that close to the nucleus the CO emission in the $A-X$ system is
self-absorbed. This follows from the observed change in the relative
intensities in various vibrational progressions as the offset from the
comet center increases (see \S~3). Self-absorption makes it difficult
to derive a reliable value for the CO column density in the absence of
a model that takes into account optical depth effects.

A detailed model of the $A-X$ system is needed to track the effects of
saturation and self-absorption. We constructed a database containing
$\sim10^{5}$ transitions between the first 50 rotational levels of each
of the 37 vibrational levels of $X^{1}\Sigma^{+}$ and the 23
vibrational levels of $A^{1}\Pi$, taking into account the energy shifts
and mixings of the transition probabilities due to interactions between
the different parity sublevels \citep{lefloch87,morton94}. Using a
simple approximation for fluorescence in subordinate lines
\citep{liu96}, expanded with a comprehensive treatment of self-absorption, our model offers an excellent fit to the data. The $A-X$ fluorescence model is used to derive spatial
profiles of the CO column density for the three comets observed by STIS.
We find that the column density profiles are consistent with a dominant
native source in all comets. The resulting production rates range
between $3.56 \times 10^{26}$ and
$1.76 \times 10^{28}$~molecules~s$^{-1}$ and are corroborated by the 
results from high
resolution $FUSE$ observations of C/2001~Q4~(NEAT) and
C/2000~WM$_{1}$~(LINEAR) \citep{weaver02}. The $HST$-GHRS observations
of comet C/1996~B2~(Hyakutake), a comet with a high CO
production rate \citep{biver99,disanti03} and thus
strongly affected by large optical depth effects, require the addition of geometrical corrections
to the simple plane-parallel atmosphere model in order to reproduce the relative line strengths.

The $HST$ observations are described in \S~2. Details about the model
and its simplifying assumptions are found in \S~3. Detailed data
analysis follows in \S~4, and a discussion of the results is given in
\S~5. We conclude with a summary in \S~6.

\section{OBSERVATIONS}

The $HST$ observations are summarized in Table~\ref{table1}. Comets
Ikeya-Zhang and C/2000~WM$_{1}$ (LINEAR) were observed near times of
high solar activity, while C/2001~Q4~(NEAT) and Hyakutake were observed
closer to solar minimum. The $HST$-STIS observations used the G140L
grating and the 52X0.2 aperture (25\arcsec~$\times$~0\farcs2 for L-mode
MAMA), resulting in a spatial resolution of $\sim$0\farcs1 and a
spectral resolution of $\sim$4~\AA\ in the wavelength range
1150--1730~\AA. The STIS instrument performance is described in
\citet{kimble98} and \citet{woodgate98}. The $HST$-GHRS \citep{heap95}
observations of comet C/1996~B2~(Hyakutake), covering the
1290--1590~\AA\ bandpass with a spectral resolution of $\sim$4~\AA, do
not provide spatial information. The GHRS Large Science Aperture (LSA)
translates into a 1\farcs74~$\times~$1\farcs74 post-COSTAR projected
area. Given that $FUSE$ and $HST$-STIS observations of comets
C/2000~WM$_{1}$~(LINEAR) and C/2001~Q4~(NEAT) were nearly simultaneous,
we choose to include in Table~\ref{table1} a summary of $FUSE$
observations for completeness \citep{feld02b,weaver02}.

\section{FLUORESCENCE MODEL FOR THE $A^{1}\Pi$~--~$X^{1}\Sigma^{+}$
SYSTEM OF CO}

The Fourth Positive Group of CO, $A^{1}\Pi$~--~$X^{1}\Sigma^{+}$, has
non-negligible overlap integrals for most of the non-diagonal
vibrational transitions. While for the $C-X$ and $B-X$ systems it is
enough to model just the (0--0) and (1--0) bands, for the $A-X$ system we must
take into account all bands connecting all 37 vibrational levels of the
$X^{1}\Sigma^{+}$ state with all 23 vibrational levels of the
$A^{1}\Pi$ state. Using only the first 50 rotational levels, which
should be sufficient for typical physical conditions in cometary comae,
the final database contains almost $10^{5}$ transitions. The latest
values for the parameters of the $A^{1}\Pi$
\citep{kurucz76,morton94,lefloch92} and $X^{1}\Sigma^{+}$ states
\citep{george94} were used to derive the energy levels and transition
wavelengths. Following \citet{morton94}, the rotational
transition probabilities $A_{J\oneprime J\arcsec}$ were obtained from
the band transition probabilities $A_{v\oneprime v''}$
\citep{beegle99,eidelsberg99,borges01,kirby89}. Even though there are
three rotational branches (P, Q and R, $\Delta J=0,\pm 1$) possible
from the same rotational level of the $A^{1}\Pi$ state ($J'$),
due to the parity selection rule, the excitation rates and branching
ratios for the P and R branches do not mix with those for the Q branch.
Each $J\oneprime$ level of the $A^{1}\Pi$ state is split into 2 opposite
parity sublevels ($\Lambda$-doubling) that have different energies and
interactions with neighboring levels of other electronic states
\citep{morton94}. The energy level shifts and the changes in the
lifetimes due to these interactions were taken into account when
available \citep{morton94,kittrell93,lefloch87}.

\begin{figure}
\begin{center}
\epsscale{0.8}
\rotatebox{90}{
\plotone{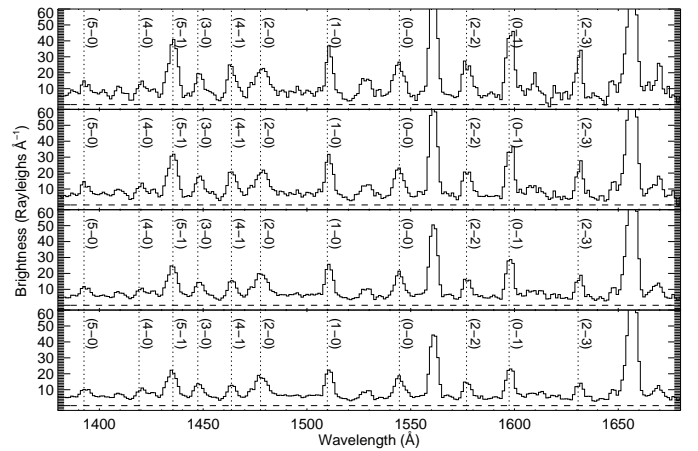} }
\caption{ Spectra of comet 153P/Ikeya-Zhang extracted from apertures that are 1, 2, 4 and 6 arcseconds wide in the spatial direction, top to bottom, centered on the comet nucleus, showing the effects of self-absorption (see text). The strong features near 1564~\AA\ and 1657~\AA\ are emissions from atomic carbon.\label{vlines}}
\end{center}
\end{figure}

In a first approximation, the fluorescent emission in the Fourth
Positive Group of CO is modeled following the prescription
from~\citet{liu96}. This model successfully accounts for saturation in the absorption of the
exciting radiation, using a Voigt profile with line overlap. A quick look at the data, however, shows that this
approach only partially accounts for optical depth effects, and we must
extend the model to include self-absorption in the fluorescent cascade.
The STIS spectra of comet 153P/Ikeya-Zhang shown in Figure~\ref{vlines}
are averaged over effective apertures of increasing size, centered on
the comet nucleus. Self-absorption is easily recognized by noting that
the bands connecting to the ground vibrational level (0--0, 1--0, 2--0,
3--0, 4--0, 5--0) show little decrease in brightness as we increase the integrated slit area and the average
column density becomes lower, while the other bands from the
progressions originating in the same upper level (see the unblended
0--1, 1--3, 2--2, 2--3, 4--1 and 5--1) decrease strongly in intensity,
so that at low column densities the relative line ratios are in
agreement with the optically thin limit.

\begin{figure}
\begin{center}
\epsscale{0.8}
\rotatebox{90}{
\plotone{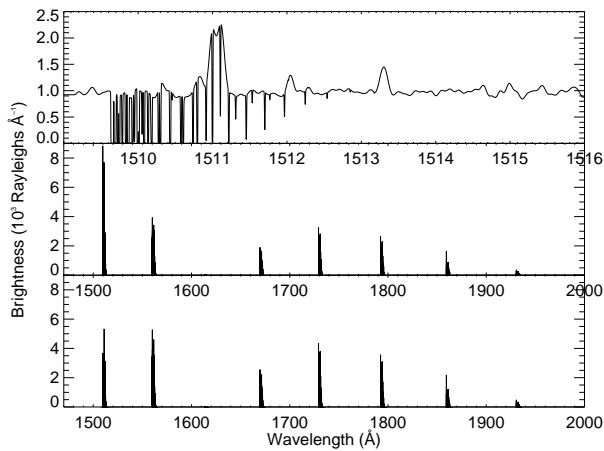} }
\caption{ Illustration of the CO fluorescence model. The absorption profile for the $A-X$ (1-0) band is applied to the solar spectrum (upper panel) and the subsequent emission in the ($1-v\arcsec$) progression is shown before and after including self-absorption (middle and lower panels, respectively).\label{mod}}
\end{center}
\end{figure}

The treatment of self-absorption is made under the assumption of local
thermodynamic equilibrium (LTE), using the fact that the photon mean
free path corresponds to an optical depth $\tau$ of one. Given that the gas temperature in the coma is less than 100~K for our observations, under LTE
conditions we need only to take into account absorption out of the
lowest (0) vibrational level of $X^{1}\Sigma^{+}$. Unlike for H$_{2}$ for example,
the ro-vibrational levels of the ground state of CO have short
lifetimes, making LTE a good approximation. The lines connecting to
v\arcsec~=~0 for which $\tau$ at line center is greater than one are
considered self-absorbed. Only an effective column density
$N_{0}^{ij}=1/\sigma_{line-center}^{ij}$, smaller than the total
absorbing column $N$, will contribute to the observed
emission in a self-absorbed transition $i-j$. The excess brightness,
due to absorption by the total column density $N$, is then redistributed
among the optically thin lines originating in the same upper level $i$
according to the branching ratios, as seen in Figure~\ref{mod}.
 
The model is constructed in the approximation of a uniform plane-parallel
atmosphere. This approximation is of limited use, due to the spherical
symmetry of the comet atmosphere and the Sun-comet-Earth geometry. The
breakdown is more apparent for larger column densities, such as in the
case of comet C/1996~B2~(Hyakutake), discussed in \S~4.2. In the
optically thin regime, valid at larger distances from the comet
nucleus, as well as for unresolved objects, such geometrical
effects are negligible. Introducing geometrical corrections in our
model can be done by allowing the column density entering the
absorption step to be different from the column density used for emission
and self-absorption. This procedure accounts for the fact that the
projected column density along the line-of-sight is larger than the
column density towards the exciting radiation source. This difference leads both
to the decrease in line saturation and to the enhancement of the
optically thin lines versus the optically thick ones due to more
self-absorption. The method is very robust, providing line intensities in agreement with the data.

\section{DATA ANALYSIS}

Although the brightness of the $A-X$ bands for the same comet should
differ slightly from one observation to another due to varying comet
heliocentric and geocentric distances, as well as due to possible
periodic variations in the volatile vaporization rate, this variation
is within the error bars of the observations and the use of an averaged
STIS spectrum for each comet in order to improve the signal-to-noise
ratio is warranted. We also select only the datasets that do not show significant deviations in background and intensity from one another.

 High resolution solar spectra from the Ultraviolet
Spectrometer Polarimeter Experiment on the {\it Solar Maximum Mission}
\citep{Tandberg-Hanssen:1981} were used, scaled to match the solar
activity at the time of comet observations as deduced from {\it
UARS}/Solstice solar flux measurements \citep{Rottman:2001}. The solar
spectrum is shifted according to the comet motion relative to the Sun
\citep[Swings effect,][]{Dymond:1989}.  For the solar \ion{H}{1} 
Lyman-$\alpha$ and -$\beta$ lines the {\it SOHO}/SUMER data of
\citet{Lemaire:2002} were used.

\begin{deluxetable*}{ccccc}
\tablecaption{Model parameters. \label{table2}}
\tablewidth{0pt}
\tablehead{
\colhead{Comet Name} & \colhead{$T_{\rm ROT}$}   & \colhead{Doppler $b$ parameter}   &
\colhead{Solar Activity} & \colhead{$N_{\rm CO}$ Range} \\ 
	& (K) & (km~s$^{-1}$) & & (10$^{14}$ cm$^{-2}$) }
\startdata
153P/Ikeya-Zhang & 82\tablenotemark{a} & 0.91\tablenotemark{b}& max & 61.3--1.49\\
C/2001 Q4 (NEAT) & 68\tablenotemark{c} & 0.79\tablenotemark{d} & min & 68.4--1.86\\
C/2000 WM$_{1}$ (LINEAR) & 77\tablenotemark{c} & 0.72\tablenotemark{b} & max & 0.935--0.327\\
C/1996 B2 (Hyakutake) & 72\tablenotemark{e} & 2.0\tablenotemark{f} & min & 145 \\
 \enddata


\tablenotetext{a}{From 74$\times r_{\rm h}^{-0.93}$~K dependence, \citet{dellorusso04}.}
\tablenotetext{b}{\citet{biver06}.}
\tablenotetext{c}{Temperature of cold component, $FUSE$ observations (see text).}
\tablenotetext{d}{Estimated from outflow velocity 0.8$\times r_{\rm h}^{-0.5}$.}
\tablenotetext{e}{From 63$\times r_{\rm h}^{-1.06}$~K dependence, \citet{disanti03}.}
\tablenotetext{f}{Within the range of values given by \citet{wouterloot98}.}
\end{deluxetable*}

The only free parameter of the model is the column density,
which is varied over a grid of values until the best fit is found. All model parameters used for each comet are summarized in
Table~\ref{table2}. The values for the rotational temperature and Doppler $b$ parameter
were chosen in agreement with infrared and radio measurements, when available. The Doppler $b$ parameter is given by $b=\sqrt{v_{therm}^{2}+v_{outflow}^{2}}\approx v_{outflow}$, where v$_{outflow}$ is the source of non-thermal line broadening, and the thermal velocities are comparable to the uncertainties in v$_{outflow}$. Molecular lines of water and other molecules are well resolved by radio observations (usually to better than 0.1~km~s$^{-1}$), and have a Gaussian profile, reflecting the symmetric outflow of the gas \citep{lecacheux03,wouterloot98,biver99}. Our observations have too low a resolution to be used to determine the Doppler line widths directly, so we rely on the published values for the expansion velocity of the gas, derived from the line profiles after correcting for thermal and instrumental broadening.

\subsection{$HST$-STIS Observations}
 
 \begin{figure}
\begin{center}
\epsscale{1.2}
\rotatebox{0}{
\plotone{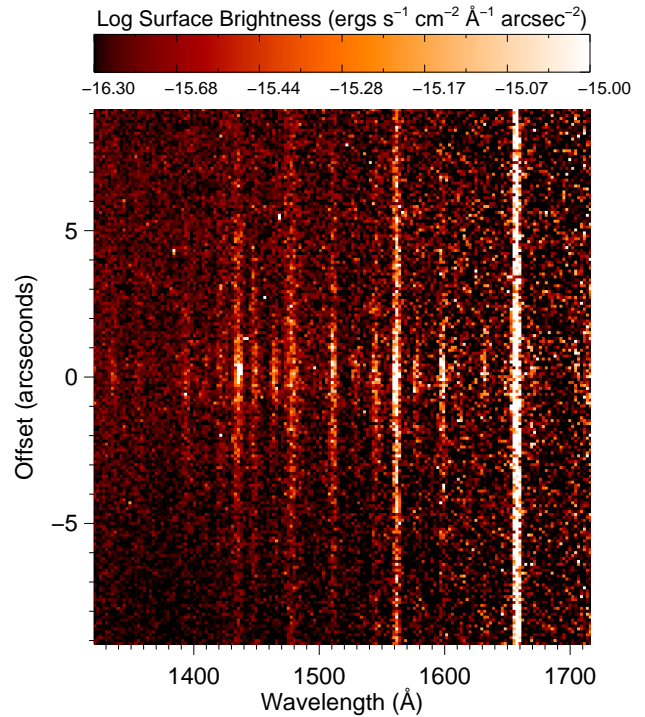} }
\caption{ $HST$-STIS spectral image of comet 153P/Ikeya-Zhang, showing the line brightness variation along the slit. The spectral region containing the strong geocoronal \ion{H}{1} $\lambda$1216 and \ion{O}{1} $\lambda$1302 has been excluded. The carbon multiplets at 1561.0~\AA\ and 1657.6~\AA\ are relatively constant due to the extended source component. The zero point in the vertical direction marks the location of the brightness peak, chosen as the center of our integration bins.   \label{stisim}}
\end{center}
\end{figure}

The high spatial resolution of the $HST$-STIS instrument is illustrated by the spectrum of comet 153P/Ikeya-Zhang in Figure~\ref{stisim}. The CO $A-X$ lines form the majority of the observed features. Their intensity decreases rapidly along the slit, in contrast with the extended \ion{C}{1}~$\lambda$1561.0 and $\lambda$1657.6 multiplets. 
Using the spatial information available, we derive the CO spatial profile for each comet by fitting spectra extracted from regions of 1\farcs5 width
at increasing offsets from the comet center. The selected regions sample areas of varying column density, for which
our model estimates an average value. The comet nucleus is identified with the center of brightness, located at zero in Figure~\ref{stisim}. The innermost region extracted, centered on
the nucleus, is noisier due to the small integrated
area. For intermediate regions the signal is better, due to the
averaging of two regions, symmetric about the comet center. For each
region, the background subtraction is performed by fitting a quadratic
polynomial to selected points from feature-free intervals. These points
were selected such that the outliers are
discarded and the resulting polynomial fit is optimal.

 The best-fit column density for
each region and its standard deviation are derived by minimizing the
$\chi^{2}$ statistics, taking into account the errors in background
subtraction.  The range of column densities obtained for each comet
observed by STIS is listed in Table~\ref{table2}, together with
the average value for comet C/1996 B2 (Hyakutake). The results are
further compared with the Haser native source model
\citep{haser57,opal77}, with an outflow velocity of
$0.85~r_{\rm h}^{-0.5}$~km~s$^{-1}$, where $r_{\rm h}$ is the comet-Sun
distance in AU \citep{budzien94,biver99}, and a CO lifetime of $2
\times 10^5$ s. We derive the CO production rate for each comet by a least squares
fit of the Haser model to the radial column density profile.  The
native source model is integrated over rectangular regions matching the
1\farcs5 spectral extractions along the STIS slit.  The resulting
production rates and their magnitude relative to water are
listed in Table~\ref{table3}.

\subsubsection{153P/Ikeya-Zhang}

\begin{figure*}
\begin{center}
\epsscale{0.75}
\rotatebox{0}{
\plotone{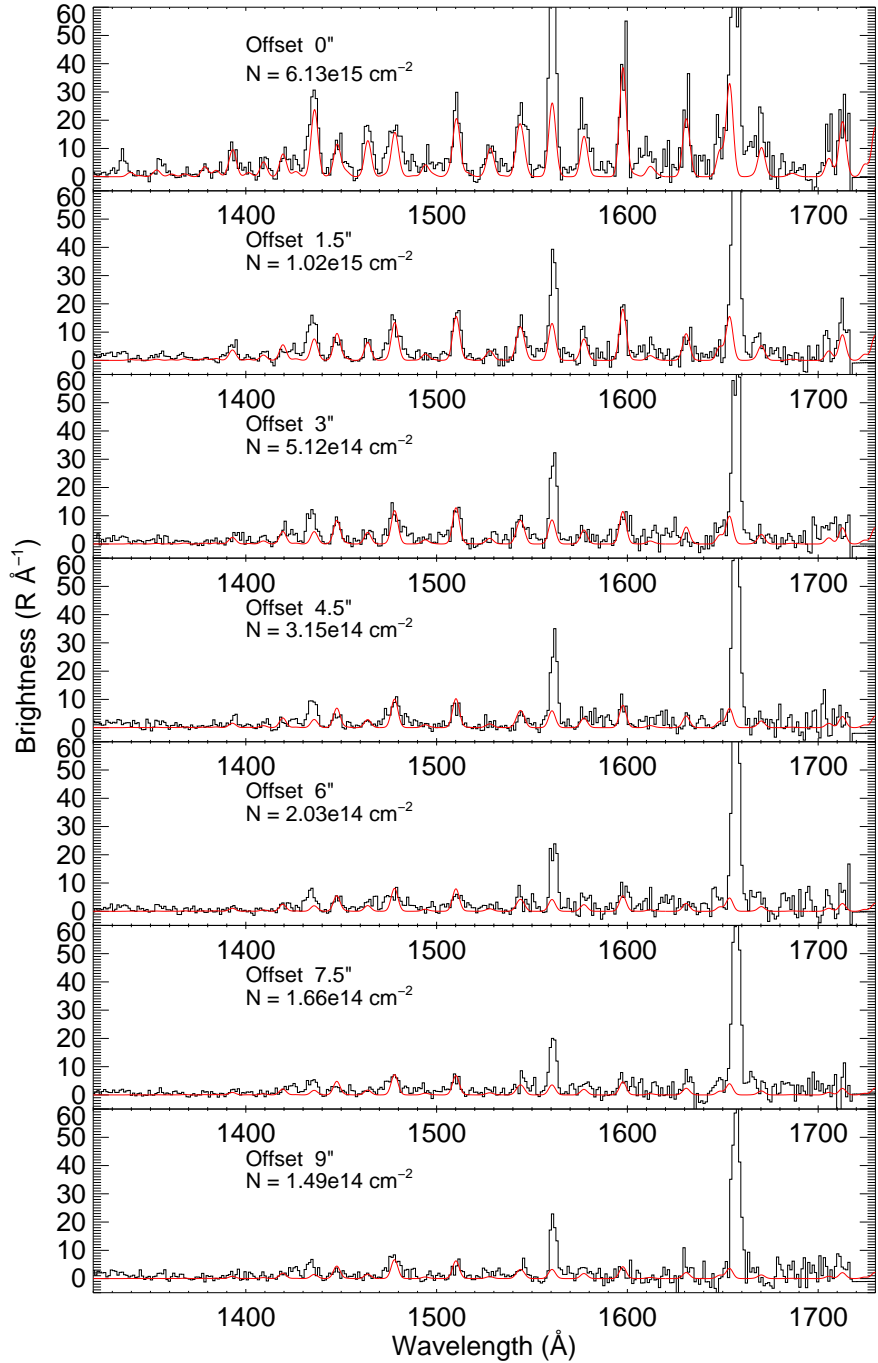}}
\caption{ $HST$-STIS spectra of comet 153P/Ikeya-Zhang extracted from 1.5\arcsec\ intervals at increasing offsets from the center of brightness, assumed to be the location of the nucleus. The red line represents the model spectrum of the CO $A-X$ system for the best fit column density for each offset. The corresponding offsets and the best fit column densities are indicated on each panel. Other emission features belong to \ion{C}{2}, \ion{C}{1}, \ion{O}{1}, \ion{S}{1}, and H$_{2}$, as indicated in Figure~\ref{h2}.\label{bigplot}}
\end{center}
\end{figure*}

\begin{figure}
\begin{center}
\epsscale{1.0}
\rotatebox{0}{
\plotone{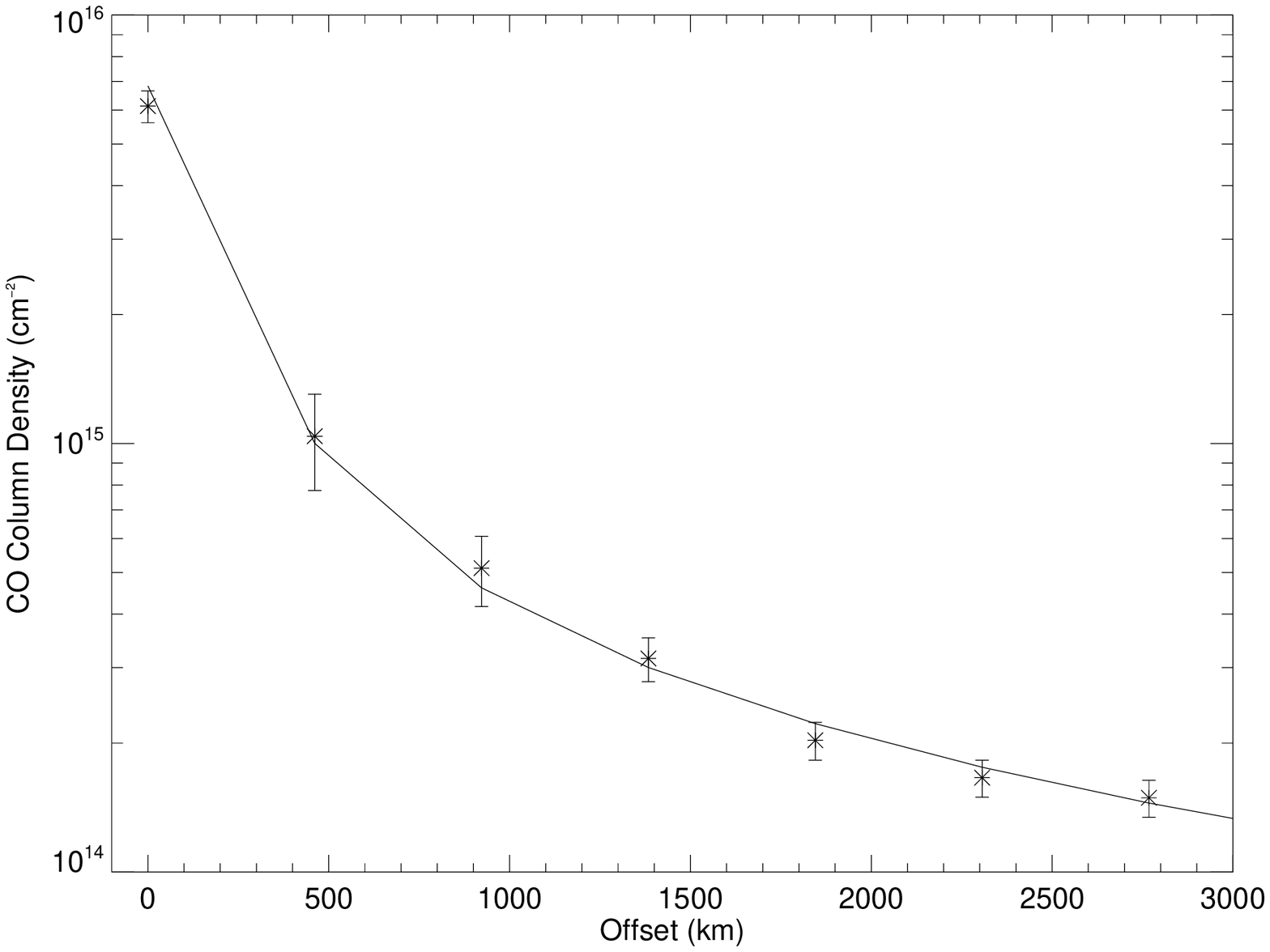} }
\caption{ Spatial profile of the CO column density for comet 153P/Ikeya-Zhang. The stars with error bars represent the column density values derived from averaged spectra using the method presented in the text. The continuous line represents the column density profile predicted by the native source model when averaged over slits covering the same areas as the observations, with a production rate of 1.54$\times$10$^{28}$~molecules~s$^{-1}$.\label{izp}}
\end{center}
\end{figure}

The values for the rotational temperature and Doppler $b$ parameter are
82~K and 0.91~km~s$^{-1}$ respectively, derived from infrared and radio
measurements~\citep{dellorusso04,biver06}. Spectra extracted from
1\farcs5 intervals were fitted using CO column densities ranging from
6.13$\times$10$^{15}$ to 1.49$\times$10$^{14}$~cm$^{-2}$, as shown in
Figure~\ref{bigplot}. The data shown are obtained by averaging the
first and third STIS observations (Table~\ref{table1}), and the best
fit model is overplotted in red. The second observation was not
included due to the background mismatch with the other two. The derived values
for the CO column density in each region are represented by stars in
Figure~\ref{izp}, plotted  as a function of the distance from the comet
nucleus. The error bars are given by the $1\sigma$ confidence level
from the $\chi^2$ statistics. Fitting to this radial profile a native
source model integrated over rectangular regions with the same coverage
as the extracted spectra~\citep{opal77}, we obtain a production rate of
$1.54\pm0.09\times10^{28}$ molecules~s$^{-1}$. The native source model
for the derived production rate is shown as a continuous line in the
same figure. The resulting CO production rate relative to water is
about $7.2\pm0.4\%$. The water production rate ($2.15\times10^{29}$
molecules~s$^{-1}$) was obtained from a vectorial model fit to an
$HST$-STIS observation of the OH (0-0) band made on 2002 April 21 at
12:19 UT.

\begin{figure}
\begin{center}
\epsscale{1.0}
\rotatebox{0}{
\plotone{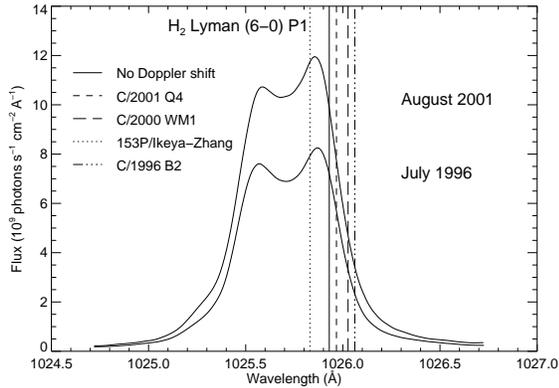}}
\caption{ Profile of solar Ly$\beta$ at maximum (August 2001) and minimum solar activity (July 1996). The velocity shifts of the absorbing H$_{2}$ (6--0) P1 line are shown for each comet.\label{Lyb}}
\end{center}
\end{figure}

\begin{figure}
\begin{center}
\epsscale{0.9}
\rotatebox{90}{
\plotone{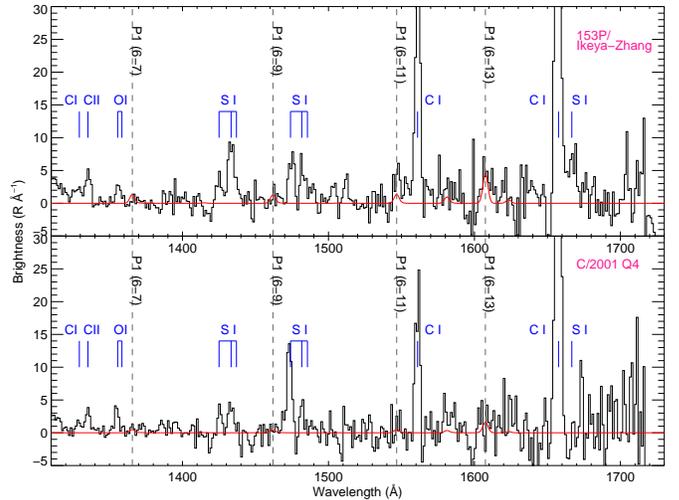}}
\caption{ Residuals from the spectrum of 153P/Ikeya-Zhang and C/2001 Q4 after subtracting the best fit CO model ($N=3.07 \times 10^{15}$~cm$^{-2}$ and $3.49 \times 10^{15}$~cm$^{-2}$, respectively). The comet spectra were extracted from a region of 4\arcsec\ width, centered on the comet nucleus. The red line is the predicted H$_2$ fluorescence spectrum pumped by solar Lyman-$\beta$ for a column density of $1.0 \times 10 ^{14}$~cm$^{-2}$ and a rotational temperature of 200~K.  Other atomic contributions are shown in blue.\label{h2}}
\end{center}
\end{figure}

The residuals after subtracting the fluorescence spectrum for the CO $A-X$ system reflect the contributions of atomic species and allow the first detection of H$_{2}$ at wavelengths longward of 1200~\AA. The H$_{2}$ spectrum consists of the P(1) lines of the Lyman (6-v\arcsec) progression pumped by solar \ion{H}{1} Lyman-$\beta$. The H$_{2}$ lines with v\arcsec = 1--3 were first detected in the {\it FUSE} spectrum of comet C/2001~A2 (LINEAR)~\citep{feld02b}. The lines at longer wavelengths, including the strongest one in the progression (v\arcsec = 13 at 1607.5~\AA), remained undetected due to the abundance of CO features and lower resolution of the STIS instrument. The solar maximum value of flux, together with a velocity shift that placed the (6-0)~P(1) line at the peak of the line, made it particularly fortuitous to detect the (6-13)~P(1) line in comet Ikeya-Zhang. Figure~\ref{Lyb} shows the shape and intensity of the solar Ly$\beta$ at minimum and maximum activity and the Doppler shifts of the H$_{2}$ absorption line corresponding to each comet. Although H$_2$ lines pumped by Ly$\beta$ were detected in comet C/2001 Q4 \citep{feld04}, due to the large negative Doppler shift and the low solar activity, in the STIS bandpass the signal-to-noise for the H$_{2}$ lines is too low to warrant a detection. A comparison of the residuals for the two comets after subtracting the CO fluorescence model is shown in Figure~\ref{h2}. The STIS spectrum was integrated over a 4\arcsec\ wide region centered on the comet nucleus, and the CO column density for the two comets ($N=3.07 \times 10^{15}$~cm$^{-2}$ and $3.49 \times 10^{15}$~cm$^{-2}$, respectively) has been obtained through the $\chi^2$ minimization procedure. The residuals for comets C/2000 WM1 and Hyakutake (not included in the figure) show no evidence for H$_{2}$, which can be explained by the large negative Doppler shifts with respect to the solar Ly$\beta$, the low outgassing rate of C/2000 WM1 and the minimum solar activity at the time of Hyakutake observations. 

The synthetic H$_{2}$ fluorescence spectra shown in red in Figure~\ref{h2} are constructed under the assumption of an H$_2$O photodissociation source for H$_{2}$. According to the dissociation model, H$_{2}$ is rotationally hot (100-300~K) and its rotational levels are in statistical equilibrium, with an ortho/para ratio of 3, similar to water \citep{budzien94,water00,bonev07}. The uncertainty in the exact value for the rotational temperature has little impact on the emerging spectrum, as the population of the J=1 level of the ground state does not differ significantly for rotational temperatures ranging from 100 to 300~K. The fluorescence efficiencies, or g-factors, used in the model were revised from those cited by \citet{feld02b}, using the solar Lyman-$\beta$ profiles obtained over the past solar cycle by \citet{Lemaire:2002}, and accounting for the comet's heliocentric velocity. The quality of the data does not warrant a $\chi^2$ fit for the H$_{2}$ fluorescence model, so we restrict ourselves to making rough estimates. Using an H$_{2}$ column density of $1.0 \times 10^{14}$~cm$^{-2}$ at a rotational temperature of 200~K we obtain a reasonable agreement with the residuals for 153P/Ikeya-Zhang (Figure~\ref{h2}, upper panel). As was the case for comet C/2001~A2, this value for the H$_2$ column is consistent, within the rather large uncertainties in both the data and the models, with an H$_2$O photodissociation source. Constraining the H$_2$ production helps our understanding of this H$_2$O dissciation channel, for which little laboratory data is available.

Under our choice of parameters, the model for 153P/Ikeya-Zhang predicts about 5.5~R for the v\arcsec = 7,9 and 11 lines, at the level of the errors due to noise and CO subtraction, leaving only the (6--13)~P(1) line detectable at a $\sim$3$\sigma$ level, with 18.4~R. This line is not detected in any other comet from our sample. For comparison, we model the H$_{2}$ fluorescence for comet C/2001 Q4. We use the same column density of H$_{2}$, as both C/2001 Q4 and 153P/Ikeya-Zhang have similar water production rates (see notes to Table~\ref{table3}). Due to the low solar activity and the unfavorable Doppler shift, the predicted line intensities are too low to be detected, consistent with the residuals (Figure~\ref{h2}, lower panel). The discrepancies between the H$_{2}$ models and the residuals are due to the uncertainties in the CO fluorescence model itself as well as to the noise in the data increasing towards longer wavelengths.

The strongest atomic lines seen in the residuals are also indicated in
Figure~\ref{h2}.  In addition to those lines usually seen in comets
\citep{mcp99}, we also note the presence of the \ion{S}{1}~($^1\!D -
^1\!D^o$) transition at 1666.7~\AA.  This transition is analogous to
\ion{O}{1}~($^1\!D - ^1\!D^o$) at 1152.2~\AA\ \citep{feld02b} and
\ion{C}{1}~($^1\!D - ^1\!P^o$) at 1930.9~\AA\ \citep{tozzi98}.
Using a g-factor of $1.92 \times 10^{-5}$ photons~atom$^{-1}$~s$^{-1}$
at 1 AU, the $\sim$30 rayleigh brightness corresponds to an average
column density of \ion{S}{1}~$^1\!D$~atoms in the aperture of $1.24
\times 10^{12}$~cm$^{-2}$.  Since the lifetime of the metastable
$^1\!D$ state of sulfur is only 28~s, this requires that within the
aperture $^1\!D$ atoms be produced at a rate of $1.9 \times
10^{26}$~s$^{-1}$ (0.09\% relative to water).  Collisional de-excitation of $^1\!D$ near the
nucleus would raise this number.  The likely source of these atoms is
the photodissociation of sulfur-bearing molecules such as H$_2$S or
CS$_2$, which must be produced at a rate greater than 0.1\% relative to
water.

\subsubsection{C/2001 Q4 (NEAT)}

\begin{figure}
\begin{center}
\epsscale{1.0}
\rotatebox{0}{
\plotone{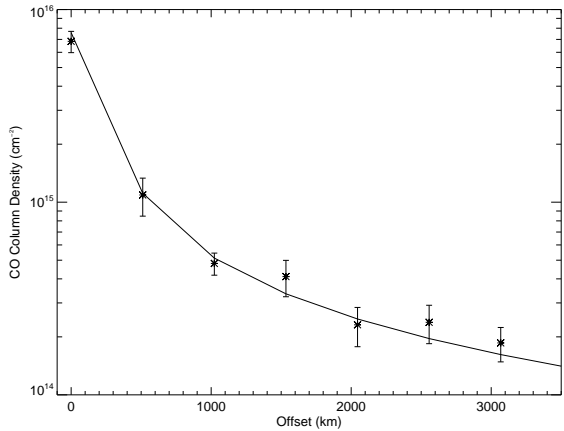} }
\caption{ Same as Figure~\ref{izp}, for comet C/2001 Q4 (NEAT). The production rate for the native source model was $1.76 \times 10^{28}$~molecules~s$^{-1}$.\label{q4p}}
\end{center}
\end{figure}

Adjusting the model parameters to the conditions of NEAT~Q4
observations we derive in the manner described for 153P/Ikeya-Zhang a CO production rate of
$1.76\pm0.16\times10^{28}$~molecules~s$^{-1}$, or $8.8\pm0.8\%$
relative to water. For the water production rate, we derive a value of
$2.0 \times 10^{29}$ molecules~s$^{-1}$ from STIS observations on
2004-04-23 21:39 UT. The CO model was fitted to the average of the
first two STIS observations (Table~\ref{table1}). For the third
observation the detector background level is much higher than for the
other two. The adopted values for the rotational temperature and
Doppler $b$ parameter are listed in Table~\ref{table2}. $FUSE$
observations of the CO $C-X$ Hopfield-Birge system at 1088~\AA\ reveal
a band profile consistent with a two component temperature model
\citep{feld02b}. The hot component ($\sim$600~K) is believed to
describe an extended CO source due to the dissociation of CO$_{2}$
\citep{feld06b}. We use for our model the temperature of the cold
component, estimated at $\sim$68~K, characteristic of the native CO
source which dominates at the smaller cometocentric distances probed by
STIS. Lacking direct radio measurements of the line widths, we use a value of 0.79~km~s$^{-1}$ for the Doppler $b$ parameter,
based on the outflow velocity 0.8$\times r_{\rm h}^{-0.5}$. The radial
profile of the best-fit column densities and the native source model
are plotted in Figure~\ref{q4p} as stars and continuous line,
respectively.

Using a two-component fit to the CO $C-X$ band observed by $FUSE$ \citep{feld02b} we derive CO column densities of $1.2 \pm 0.3 \times 10^{14}$~cm$^{-2}$ and $2.38 \pm 0.08 \times 10^{13}$~cm$^{-2}$ for the cold and hot component, respectively, averaged over the entire 30\arcsec~$\times$~30\arcsec slit. A native source for the cold component requires a production rate of $1.36\pm0.40\times10^{28}$~molecules~s$^{-1}$, or $6.8\pm2.0\%$ relative to water. The solar flux pumping the $C-X$ fluorescence was based on quiet sun whole disk data from {\it SOHO}/SUMER \citep{Curdt:2001}, normalized, at wavelengths longward of 1200 \AA, to {\it UARS}/SOLSTICE solar flux measurements appropriate to the solar activity at the time of our observation \citep{Rottman:2001}.

\subsubsection{C/2000 WM$_{1}$ (LINEAR)}

\begin{figure}
\begin{center}
\epsscale{1.0}
\rotatebox{0}{
\plotone{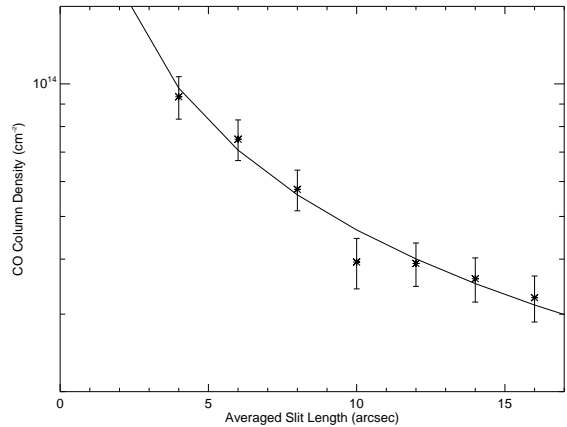}}
\caption{ CO column density profile for comet C/2000 WM$_{1}$ (LINEAR) extracted from regions of increasing widths, centered on the comet nucleus. The continuous line corresponds to the values predicted by the native source model with a production rate of $3.56 \times 10^{26}$~molecules~s$^{-1}$, when integrated over similar slit sizes.\label{wm1p}}
\end{center}
\end{figure}

The CO emission detected by STIS in the observation of comet C/2000
WM$_{1}$ (LINEAR) was too weak to allow us to repeat the same analysis
performed in the case of comets 153P/Ikeya-Zhang and C/2001 Q4 (NEAT).
Instead, we chose to integrate the STIS spectrum over increasing widths
centered on the nucleus, in order to make use of the stronger signal in
the center and to increase the number of contributing pixels. We
started with a 4\arcsec\ wide region which was increased progressively
up to 16\arcsec. For the rotational temperature we used the 77~K value
derived for the cold component using $FUSE$ observations
\citep{weaver02}, while the Doppler $b$ parameter value of
0.72~km~s$^{-1}$ was chosen to match the radio observations of
\citet{biver06}. All three STIS observations were averaged together to
obtain detectable CO emission features. The best-fit column densities
over the selected regions are plotted in Figure~\ref{wm1p} as a function of
the integrated slit width. Fitting a native source model integrated
over the same rectangular regions we obtain a CO production rate of
$3.56\pm0.2\times10^{26}$~molecules~s$^{-1}$. This model is shown by a
continuous line in Figure~\ref{wm1p}. A CO production rate of
$0.44 \pm 0.03$\% relative to water is obtained adopting the favored
H$_{2}$O production rate for the $FUSE$ observations 
\citep[8.0$\times$10$^{28}$~molecules~s$^{-1}$,][]{weaver02}.
This makes C/2000 WM$_{1}$ (LINEAR) the most CO-poor comet of our sample.

\begin{deluxetable*}{cccc}
\tablecaption{Production rates. \label{table3}}
\tablewidth{0pt}
\tablehead{
\colhead{Comet Name} & \colhead{$Q_{\rm CO}$\tablenotemark{$\star$}}   & \colhead{$Q_{\rm CO}$/$Q_{\rm H_{2}O}$} & {Other $Q_{\rm CO}$ Measurements}\\ 
	& (10$^{28}$~molecules~s$^{-1}$) & (\%) & (10$^{28}$~molecules~s$^{-1}$)
}
\startdata
153P/Ikeya-Zhang & 1.54 $\pm$ 0.09 & 7.2 $\pm$ 0.4\tablenotemark{a} & 0.73 $\pm$ 0.16\tablenotemark{b}\\
C/2001 Q4 (NEAT) & 1.76 $\pm$ 0.16 & 8.8 $\pm$ 0.8\tablenotemark{c} & 1.36 $\pm$ 0.40\tablenotemark{d}\\
C/2000 WM$_{1}$ (LINEAR) & 0.036 $\pm$ 0.002 & 0.44 $\pm$ 0.03\tablenotemark{e} & 0.035 $\pm$ 0.003\tablenotemark{f}\\
C/1996 B2 (Hyakutake) & 4.97 $\pm$ 0.07 & 20.9 $\pm$ 0.3\tablenotemark{g} & 4.84 $\pm$ 0.58\tablenotemark{h}\\
 \enddata


\tablenotetext{$\star$}{The error bars are given by the 1 $\sigma$ interval from the $\chi^{2}$ statistics. In addition, we estimate that systematics amount to a 15\% uncertainty in the production rates for the STIS observations.}
\tablenotetext{a}{Water production rate 2.15$\times$10$^{29}$~molecules~s$^{-1}$ from $HST$-STIS observations (see text).}
\tablenotetext{b}{Using r$_{h}^{-2.1}$ scaling from \citet{biver06}.}
\tablenotetext{c}{Water production rate 2.0$\times$10$^{29}$~molecules~s$^{-1}$ from $HST$-STIS observations (see text).}
\tablenotetext{d}{Derived in this paper from $FUSE$ observations (\S 4.1.2). The cited value includes only the cold source component.}
\tablenotetext{e}{Water production rate 8.0$\times$10$^{28}$~molecules~s$^{-1}$ used for the $FUSE$ observations \citep{weaver02}.}
\tablenotetext{f}{Based on $FUSE$ observations \citep{weaver02}, revised for this paper. The cited value includes only the cold source component.}
\tablenotetext{g}{Water production rate 2.38$\times$10$^{29}$~molecules~s$^{-1}$ from \citet{combi98}.}
\tablenotetext{h}{Value measured on April 1.2 using JCMT radio telescope \citep{biver99}. From the 4.7$\times$10$^{28}r_{\rm h}^{-2.1}$~molecules~s$^{-1}$ dependence the predicted value is 6.07$\times$10$^{28}$~molecules~s$^{-1}$.}
\end{deluxetable*}

\subsection{$HST$-GHRS Observations: C/1996 B2 (Hyakutake)}

Since the $HST$-GHRS observations do not provide spatial information,
we are unable to derive a CO column density profile for comet C/1996 B2
(Hyakutake). The CO production rate is obtained by comparing the CO
column density derived by fitting our
fluorescence model to the GHRS spectrum with the CO column density predicted for the GHRS aperture by the native
source model. As model parameters for the synthetic spectrum we used a
rotational temperature of 72~K given by the 63$\times r_{\rm h}^{-1.06}$~K
dependence from \citet{disanti03}, which is similar to the value given
by \citet{lis97}, and a $b$ parameter of 2.0~km~s$^{-1}$, within the
range of outflow velocities measured by \citet{wouterloot98} but slighlty higher than derived from the optical line widths of \citet{combi99}. A lower $b$ value is inconsistent with the total amount of absorbed solar
radiation (from the conservation of the number of photons), suggesting
larger turbulent motions in the 1\farcs74$\times$1\farcs74 area probed
by GHRS.

\begin{figure}
\begin{center}
\epsscale{0.9}
\rotatebox{90}{
\plotone{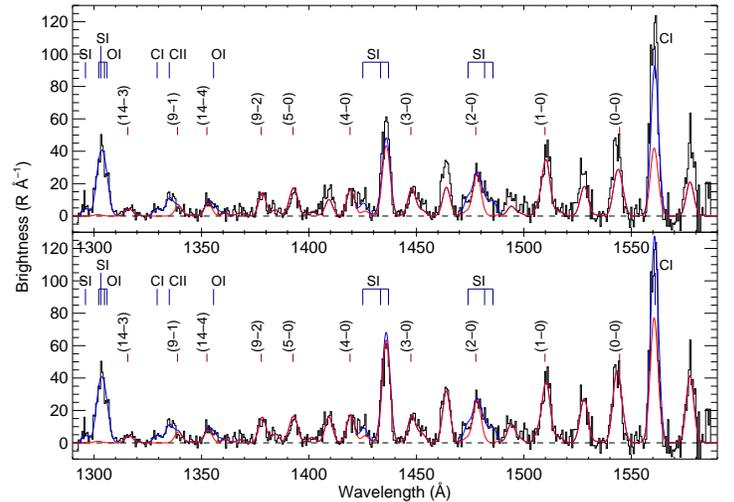}}
\caption{ $HST$-GHRS spectrum of comet C/1996 B2 (Hyakutake), and the model spectrum of the CO $A-X$ system for a line-of-sight column density of $1.45\times10^{16}$~cm$^{-2}$, before (upper panel) and after geometric corrections (lower panel).  The CO model spectrum is shown in red. The blue line contains contributions from atomic lines of \ion{C}{2}, \ion{C}{1}, \ion{O}{1}, and \ion{S}{1}, as indicated. The $A-X$ bands that do not change significantly due to the geometric effects are labeled.\label{hkt}}
\end{center}
\end{figure}

Using the CO $A-X$ fluorescence model with self-absorption we obtain a first estimate
for the CO column density averaged over the GHRS slit. However, the
predicted line ratios are not in agreement with the data (see the red
line in Figure~\ref{hkt}, upper panel). In order to improve the fit and
better constrain the column density we adjust the model to account for
the geometry of the Sun-comet-Earth system. Starting with the previous
estimate on the CO column we can constrain the ratio between the
line-of-sight column density and the absorbing column on the comet-Sun
direction. Iterating this procedure we obtain a best fit value for the
line-of-sight CO column density of
$1.45 \pm 0.87 \times 10^{16}$~cm$^{-2}$. This model is shown with a
red line in the lower panel of Figure~\ref{hkt}. The blue model in the
same figure contains contributions from atomic species
\ion{C}{2}, \ion{C}{1}, \ion{O}{1}, and \ion{S}{1}, as labeled,
which account for the remaining features in the spectrum. Under the
assumption of a native source model with the same lifetime and gas outflow velocity as employed for the STIS observations, we obtain a CO
production rate of $4.97 \pm 0.07 \times 10^{28}$~molecules~s$^{-1}$. This
represents the highest CO production rate relative to water
from our sample, $20.9 \pm 0.3$\%, using the water production rate of
2.38$\times$10$^{29}$~molecules~s$^{-1}$ from \citet{combi98}.

Figure~\ref{hkt} also clearly shows the presence of several bands
originating on the v$'=9$ and 14 levels that are pumped by solar
\ion{O}{1} $\lambda$1302 and \ion{H}{1} Lyman-$\alpha$, respectively
\citep{Wolven:1998}.  \citet{Kassal:1976} first pointed out that
scattering of Lyman-$\alpha$ by the (14,0) band is comparable to, if
not larger than, the direct solar scattering by all other bands of the
CO Fourth Positive system for CO column densities $\ge
10^{17}$~cm$^{-2}$.  The (14,4) band was subsequently identified in
the spectrum of Venus \citep{Durrance:1981}.  These bands are not detected
in any of the STIS comet spectra.  The high column density in the
field-of-view of comet Hyakutake makes them visible, albeit at low S/N,
and allows for a determination of column density independent of optical
saturation effects.  In evaluating the fluorescence efficiency of these
bands, the overlap between the solar lines and the individual lines of
the CO bands is a sensitive function of rotational temperature and
heliocentric velocity and requires accurate profiles of the solar lines
\citep{Lemaire:2002}.  The (14-3), (14-4), and (9-2) bands are included 
in the model and are seen to be completely consistent with the CO column
density derived above.

We note that the optically thick bands connecting
to the v\arcsec~=~0 level of the ground state, namely (1-0) at 1509.8
\AA, (2-0) at 1477.6 \AA, (3-0) at 1447.4 \AA, (4-0) at 1419.1 \AA\ and
(5-0) at 1392.6 \AA, all labeled in Figure~\ref{hkt}, do not show a significant variation due to
geometric corrections. This can be understood from the fact that while
the correction adjusts the line-of-sight column density relative to the
column absorbing the solar radiation, the emission in the optically
thick lines will still be determined by the column corresponding to one
optical depth, which depends only on temperature and $b$
parameter. The (0-0) band at 1544.5 \AA\ does not seem to follow this
pattern due to blending with the (3-2) band at
1542.5 \AA. The same lack of variation is exhibited by the bands
belonging to optically thin progressions pumped by solar emission
lines, such as (14-3) at 1315.7 \AA\ and (9-2) at 1377.8 \AA. This is a result
of the simple linear scaling in the optically thin limit, which makes the absorbing column indistinguishable from the emitting column. 

\section{DISCUSSION}

\subsection{Sources of Uncertainty and Comparison with Other Measurements}

The derived column densities are subject to uncertainties due to the
choice of model parameters and background subtraction. A more complete
model would involve a multidimensional $\chi^{2}$ minimization to
constrain simultaneously the column density, rotational temperature and
Doppler $b$ parameter. Aside from the fact that this approach requires
rather large computational resources, we expect that given the quality
of the data the resulting $\chi^{2}$ 3D surfaces will have rather low
contrast minima and the improvement in the resulting production rates
would be negligible. More of a concern is the background subtraction in
the $HST$-STIS data. The background is variable both in the spatial and
spectral directions from one observation to another, making it
impossible to give a comprehensive subtraction prescription. The
optimal background subtraction is determined on a case-by-case basis.
While the background-related uncertainties can lead to $\sim$30\%
variations in the values for the column densities, we estimate that the
change in the resulting production rates is only about 6\%. These values are only slightly larger than the $1\sigma$ error bars from the $\chi^{2}$ minimization. Similarly, increasing the rotational temperature from 70~K to 100~K results in a $\sim$30\% decrease in production rate. However, the rotational temperatures relevant for our observations were derived from either $FUSE$ observations or from the T$_{rot}$ vs. r$_{h}$ dependences obtained by radio measurements, and are constrained to better than $\pm$6~K. This results only in a $\sim$7\% variation in production rate, as the heliocentric distance of the comet does not vary significantly during our observations. The absolute values of the CO production rate and column density are also sensitive to the STIS calibration pipeline, which
is based on point-source stellar standards. The column densities derived from STIS data could be overpredicted by at most 30\% due to calibration offsets. Variations in the parameters used for the native source model, such as the CO lifetime and outlow velocity, could also change the production rate by a few percent. Other less quantifiable uncertainties to which the fluorescence model is particularly sensitive are the oscillator strengths and the UV solar flux, especially due to the variable emission features at
high solar activity. Overall, we estimate that the systematic errors in the production rates amount to at most 15\%.

To assess the effects of the error sources mentioned above, it is
useful to compare our results to other measurements of the CO production rate in these comets from
different spectral regions. The values obtained for comets
C/2000 WM$_{1}$ (LINEAR) and C/1996 B2 (Hyakutake) are in excellent
agreement with previous measurements. For comet C/1996 B2 (Hyakutake)
at similar heliocentric distances \citet{biver99} find vales of
$4.98 \pm 0.09 \times 10^{28}$~molecules~s$^{-1}$ (0.952~AU) and
$4.84 \pm 0.58 \times 10^{28}$~molecules~s$^{-1}$ (0.894~AU). The
production rate relative to water is comparable to previously measured mixing ratios of
14 to 19\% \citep{disanti03} and 22\% \citep{biver99}. For comet C/2000 WM$_{1}$
(LINEAR) both $FUSE$ \citep{weaver02} and radio \citep{biver06}
observations are consistent with a CO mixing ratio of $\sim$0.4\% and
a CO production rate of $\sim$3 to 4$\times$10$^{26}$~molecules~s$^{-1}$. A good agreement is again found when comparing the CO production rate derived for comet C/2001
Q4 (NEAT) with the results from $FUSE$ observations ($\S$ 4.1.2),
listed in Table~\ref{table3}. We note that the measurements based on $FUSE$ observations
used only the cold component of CO (see $\S$~4.1.2) in estimating the CO
production rate. The cold component is believed to reflect the native source of CO, which is directly probed by STIS. The two values agree marginally within the error bars, and the mismatch can be attributed to short time variability and pointing instability for the $FUSE$ observations. The count rates for the $C-X$ band in the $FUSE$ data show factor of two variations with a periodicity of $\sim$20 hours. The $FUSE$ observation overlaps with a much shorter STIS exposure, but the exact correlation in comet activity between the two datasets is hard to assess due to the different fields of view of the two instruments.

For comet 153P/Ikeya-Zhang
the derived CO production rate
is about a factor of 2 higher than the estimate from the range of
values given by \citet{biver06} (using the
$r_{\rm h}^{-2.1}$ scaling, see Table~\ref{table3}) and \citet{disanti02}. The factor of two difference may be attributed to the uncertainties in the model parameters
(rotational temperature and Doppler parameter) and in the
background subtraction. However, as discussed above, the combined sources of error should result in less than 15\% uncertainty. On the other hand, the $r_{\rm h}^{-2.1}$ dependence was derived from two measurements, one at 0.51~AU and the other at 1.26~AU, not excluding the possibility of temporal variations affecting our observations at 0.89~AU. Moreover, the uncertainties in the $r_{\rm h}^{-2.1}$ scaling have not been quantified, making our value for the production rate more reliable. The CO abundance relative to water is again higher than other estimates. However, using the water production rate given by the 19.0$\times$10$^{28}r_{\rm h}^{-4.}$~molecules~s$^{-1}$
dependence derived by \citet{biver06} from H$_2$O observations made by the
{\it Odin} satellite \citep{lecacheux03}, the CO mixing ratio
decreases from $7.2\pm0.4\%$ to $5.1\pm0.3\%$. This would place the CO
relative abundance in comet 153P/Ikeya-Zhang closer to the value of
$3.8\pm0.6\%$ at $\sim$1~AU \citep{biver06} and $4.7\pm0.8\%$ at
0.78~AU \citep{disanti02}.

\subsection{Application to Broad-band Imaging}

\begin{figure}
\begin{center}
\epsscale{1.0}
\rotatebox{0}{
\plotone{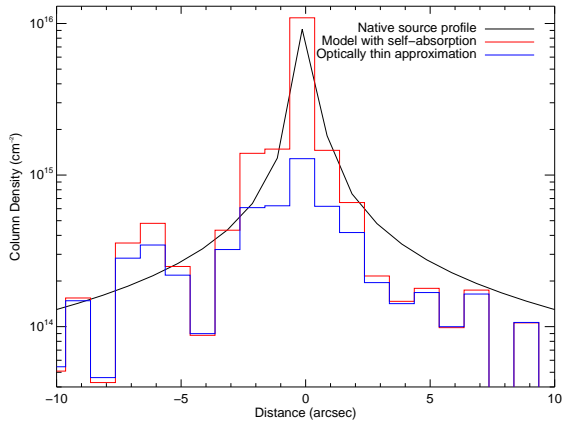} }
\caption{ Spatial profiles of the CO column density for comet 153P/Ikeya-Zhang, derived from the integrated brightness along the slit in 1\arcsec\ bins, selecting the unblended lines from the $A-X$ system. The red line uses the total brightness-column density dependence associated with the present model, while the blue line is obtained in the optically thin approximation. The continuous black line is the native source model for a production rate of 1.54$\times$10$^{28}$~molecules~s$^{-1}$, integrated in 1\arcsec\ bins.\label{izot}}
\end{center}
\end{figure}

The advantage of long-slit spectroscopy with STIS resides in giving us a better understanding of the spatial distribution of CO and the effects of increasing optical depths on the observed line ratios. We show how this information is applicable to broad-band imaging by integrating the
brightness of the $A-X$ bands in the STIS bandpass that are not contaminated by atomic lines. We then integrate the fluorescence
model over the same bands deriving an equivalent g-factor (as a ratio
between the total brightness and the column density). Given the
specific optical depth corrections in our model, the resulting g-factor
will be in fact a brightness-column density dependence rather than a
globally constant ratio. The radial brightness profile of the integrated
bands observed by STIS can be then converted into a
column density profile. This procedure can be visualized in
Figure~\ref{izot}, where the data for comet 153P/Ikeya-Zhang have been
used. The brightness profile has been rebinned in 1\arcsec\ bins and
background subtracted. The red histogram represents the column density
profile obtained using the optically thick brightness-wavelength
dependence, while the blue histogram represents the column density
profile obtained in the optically thin approximation, using a constant
g-factor. The black line is the native source model integrated in
1\arcsec\ $\times$~0\farcs2 bins, for a production rate of
1.54$\times$10$^{28}$~molecules~s$^{-1}$, as derived in \S~4.1.1. This
figure illustrates the importance of including optical depth effects in
order to predict the correct column density. This method has
been used to interpret the imaging data from the {\it HST} Solar Blind
Channel of the Advanced Camera for Surveys (ACS/SBC),
and give an estimate of the number of molecules released from
comet 9P/Tempel~1 as a result of the {\it Deep Impact} encounter
\citep{feld06a}. The uncertainties in this type of measurements come
mainly from the additional emission features of atomic lines in the
bandpass.

\subsection{The Native Source of CO}

The $HST$-STIS observations allow us to constrain the dominant source
of CO in the region probed by the 25\arcsec\ $\times$~0\farcs2 slit.
Reconstructing the surface column density distribution from the radial
profile we estimate that in a 10\arcsec\ $\times$~10\arcsec\ box the native
source contribution to the total number of molecules is as high as 80\%
for comet C/2000 WM$_{1}$ (LINEAR) and from 90\% to 99\% for comets
C/2001 Q4 (NEAT) and 153P/Ikeya-Zhang, respectively. This result suggests that the native source dominates in the inner $\sim 3000$~km from the comet center, while the extended component becomes important at larger distances. However, the native CO observed in the coma is linked to the amount of CO in the nucleus through the outgassing mechanism, which is not well known. Laboratory data suggests that the abundance of CO relative to water in the ice can be a factor of 5 to 10 lower than the one observed in the coma \citep{notesco97,colangeliii}. The exact value depends on the temperature at which the comet is outgassing.

The large variation in the CO production rate relative to water can be understood from the large range of heliocentric distances over which the comets have formed and then migrated to the outer parts of the Solar System through gravitational interactions with the planets. The observed diversity among comets is thought to be related to the local gas and dust composition and temperature where they formed. Under the assumption that CO has been trapped by water ice during comet formation, the local temperature can be estimated from the observed production rate \citep{notesco97}. The formation temperatures in our sample are expected to range from 50~K or less for comets C/2001 Q4, 153P/Ikeya-Zhang and Hyakutake, to more than 60~K for C/2000 WM$_{1}$. If the gas trapping mechanism is more efficient than deposition, as suggested by laboratory studies of CO and water interactions \citep{collings03}, important enrichments for the species with a higher sublimation temperature can occur \citep{notesco97}. This can explain the lack of correlation between the abundance of CO and that of other molecules with a similar sublimation temperature \citep{gibb03,biver02}.

\section{SUMMARY}

We have developed a fluorescence model for the interpretation of CO
$A-X$ emission observed in several recent comets by the {\it
Hubble Space Telescope} employing the latest values for the transition
wavelengths and oscillator strengths. The radiative transfer
approximation takes into account saturation effects using Voigt
profiles for each of the $\sim$10$^{5}$ transitions. Self-absorption is
introduced using a photon mean free path approximation. This process is
significant for column densities above few$\times$10$^{14}$~cm$^{-2}$, encountered close to the comet
nucleus. It is shown that the model reproduces the optically thin
limit for lower column densities, at larger distances from the
comet center. When the column densities are of the order
$\sim$10$^{16}$~cm$^{-2}$ or above, a good fit to the data requires a
distinction between the absorbing CO column in the Sun-comet direction
and the emitting column along the line-of-sight.  The approximations in
the radiative transfer model are justified for the quality of
the available data. Constraining better the optical depth effects demands an
exact treatment, which would be warranted at higher spectral resolution
in order to predict the relative strengths of individual lines
contained in the $A-X$ bands.

For the comets observed by STIS the fit of the column density profile
is consistent with a predominantly native source, with production rates
ranging from $3.56 \times 10^{26}$~to
$1.76 \times 10^{28}$~molecules s$^{-1}$. The quality and the spatial
extent of the STIS data does not allow for a detection of a small
extended source component.  We find large variations in the CO abundance
relative to water, from 0.4\% for C/2000 WM$_{1}$ (LINEAR) to 21\% for
C/1996 B2 (Hyakutake). This diversity among comets is still not well understood, as no clear trend emerges at the current stage of comet observations \citep{cometsii}. In spite of the caveats discussed in \S~5, the absolute values of the
CO production rate and column density are better constrained by the
introduction of optical depth effects, and give reasonable confidence
levels for the CO production rate and a good agreement with previous
results. Moreover, the present model proves to be a valuable tool in
analyzing broadband imaging data, such as the ACS/SBC observations of
comet 9P/Tempel~1 \citep{feld06a}.

\acknowledgements

This work is based on observations with the NASA/ESA {\it Hubble
Space Telescope} obtained at the Space Telescope Science Institute,
which is operated by the Association of Universities for Research in
Astronomy (AURA), Inc., under NASA contract NAS 5-26555.  We wish to
thank Stephan McCandliss for providing us with the H$_{2}$ UV
fluorescence model.  PDF wishes to acknowledge the hospitality of
Arcetri Observatory during a sabbatical visit in April-May 2003.  This
work was supported by NASA grants HST-GO-09185.04-A, HST-GO-09496.04-A,
and HST-GO-09906.04-A to The Johns Hopkins University.


\end{document}